\def\hetetwo{{\it HETE-2}\ }
\def\hetetwonosp{{\it HETE-2}}

\def\bsax{{\it Beppo}SAX\ }

\def\eop{E^{\rm obs}_{\rm peak}}
\def\eiso{E_{\rm iso}}

\def\ep{E_{\rm peak}}
\def\egi{E^{\rm inf}_{\gamma}}
\def\egt{E^{\rm true}_{\gamma}}

\def\thetav{\theta_{\rm v}}

\def\thetaj{\theta_{0}}
\def\egamma{E_{\gamma}}
\def\epei{\ep \propto \eiso^{1/2}}
\def\epeioff{\ep \propto \eiso^{1/3}}
      [1999/12/01 v1.4c Il Nuovo Cimento]
\documentclass{cimento}

\usepackage{graphicx}
\title{The Importance of Off-Axis Beaming in Jet Models}
\author{T. Q.~Donaghy\from{uofc}}
\instlist{\inst{uofc} Dept. of Astronomy \& Astrophysics, 
University of Chicago\\ 5640 S. Ellis Ave., Chicago, IL, 60615}
\PACSes{\PACSit{98.70.Rz}{gamma-ray sources; gamma-ray bursts}}
\newcommand{\beq}{\begin{equation}}
\newcommand{\eeq}{\end{equation}}
\begin{document}

\maketitle

\begin{abstract}
Gamma-Ray Bursts (GRBs) are widely thought to originate from collimated
jets of material moving at relativistic velocities.  Emission from such
a jet should be visible even when viewed from outside the angle of
collimation.  Using Monte Carlo population synthesis methods and
including the effects of this off-axis beaming, we can compare various
GRB jet models against the global properties of observed bursts.  We
explore whether or not the X-Ray Flashes (XRFs) seen by \hetetwo and
\bsax can be explained as classical GRBs viewed off-axis, and begin to
address the more general question of the importance of off-axis beaming
in current burst samples.

\end{abstract}

\vspace{-0.3truein}
\section{Introduction}

The importance of collimated jets in GRBs was highlighted by the
extremely large isotropic-equivalent energies of very bright events
like GRB 990123 and by the observation of breaks in afterglow
light-curves.  \cite{frail2001} and \cite{bloom2003} corrected the
isotropic-equivalent energies by the beaming fraction obtained from
afterglow light-curves and found that the values of $\egamma$ were
clustered around $10^{51}$ ergs (although see \cite{ghirlanda2004a}). 
Recent results from \hetetwo \cite{sakamoto2004b} have shown that XRFs
\cite{heise2000,kippen2002}, X-Ray Rich GRBs and GRBs lie along a
continuum of properties and that XRFs with known redshift extend the
$\epei$ relation predicted by \cite{lloyd-ronning2000} and found by 
\cite{amati2002} to over 5 orders of magnitude in $\eiso$
\cite{lamb2005}.

Relativistic kinematics implies that even a ``top-hat''-shaped jet
will be visible when viewed outside the angle of collimation, $\thetaj$
\cite{ioka2001}.  \cite{yamazaki2002,yamazaki2003} used this fact to
construct a model where XRFs are simply classical GRBs viewed at an
angle $\thetav > \thetaj$.  The authors showed that such a model could
reproduce many of the observed characteristics of XRFs. 
\cite{yamazaki2004} showed that in such a model, the distribution of
both on- and off-axis observed bursts was roughly consistent with the
$\epei$ relation.

In this paper, we explore further the possibility that the XRFs
observed by \hetetwo and \bsax are primarily off-axis GRBs.  Using and
extending the population synthesis techniques presented by
\cite{ldg2005} and \cite{dlg2005}, we present predictions for the
global properties of bursts localized by \hetetwonosp.  We show that it
is difficult to account for the observed properties of XRFs by
modelling them solely as regular GRBs viewed off-axis.  However, since
off-axis emission must exist solely on physical grounds, we seek to
understand its relative importance in large burst populations.  We
revisit the model put forward in \cite{ldg2005}, now including the
effects of off-axis beaming.  We note that rough constraints on the
bulk $\gamma$ might be found by considering the fraction of bursts that
are not consistent with the $\epei$ relation.  Finally, we consider
some possible extensions to our model.

\section{Simulations}

\cite{yamazaki2002,yamazaki2003,yamazaki2004} work with a fairly
detailed model of the burst emission; for this work, we adopt a
simpler model of off-axis beaming in GRB jets.  We make no assumptions
about the underlying physics generating the burst, and we make the
approximation that the bulk of the emission comes directly from the
edge of the jet closest to the viewing angle line-of-sight (i.e. we
ignore all integrals over the surface of the jet and time-of-flight
effects).   Our model focuses on the kinematic transformations of two
important burst quantities, $\eiso$ and $\ep$, as a function of viewing
angle.

Frequencies in the rest frame of the burst material will appear
Doppler shifted by a factor, $\delta = \gamma (1-\beta \cos\theta)$,
where $\beta$ is the velocity of the bulk material and $\theta$ is the
angle between the direction of motion and the source frame observer. 
The quantities $\ep$ and $\eiso$ then transform as $\ep \propto
\ep^{\prime} \delta^{-1}$ and $\eiso \propto \eiso^{\prime}
\delta^{-3}$.  For a burst viewed off-axis, these relations imply
$\epeioff$.  \cite{yamazaki2004} do not consider $\eiso$ to be fully
bolometric and so derive a slightly different prescription for the
off-axis relation.

We adopt an effective angular distribution of the emissivity,
$\epsilon(\thetav)$, that is uniform for $\thetav < \thetaj$ and
decreases for $\thetav > \thetaj$:
\beq
\epsilon(\thetav) = \frac{\eiso}{4\pi} 
	= \left\{ \begin{array}{l}
		A \\
		A \cdot \left( \delta/\delta_{0} \right)^{-3} \\
	\end{array} \right.   \;\;{\rm and}\;\;\;\;
\ep = \left\{ \begin{array}{ll}
		B & 
		{\rm if\;\;} \thetav \le \thetaj\\
		B \cdot \left( \delta/\delta_{0} \right)^{-1} & 
		{\rm if\;\;} \thetav > \thetaj,\\
	\end{array} \right.
\eeq
where in this expression, $\delta =
\gamma[1-\beta\cos(\thetav-\thetaj)]$, $\delta_{0} = \gamma(1-\beta)$
is the value of $\delta$ when $\thetav=\thetaj$, $A$ is a normalization
constant described below, $B = C_{\rm A} \cdot (\eiso/10^{52} {\rm
\;ergs})^{1/2}$, and $C_{\rm A}$ is drawn from a narrow lognormal
distribution.   Hence, $\ep$ obeys the $\epei$ relation inside the jet
and $\epeioff$ outside the jet.  We then define the ``true'' standard
energy by integrating this emissivity over the entire sphere:
\beq
\egt = 2 \cdot 2\pi \int_{0}^{\pi/2}
	\epsilon(\thetav) \;\sin\thetav \;d\thetav
	= 4\pi A \left[ 1-\cos\thetaj + I(\gamma, \thetaj) \right].
\eeq

We define the $\egamma$ value inferred via the method of
\cite{frail2001} to be $\egi = 4\pi A ( 1-\cos\thetaj )$.  The presence
of beaming implies $\egi \ne \egt$.  We fix our normalization constant,
$A$, to match the original prescription used by \cite{frail2001}, by
drawing values for $\egi$ from a narrow lognormal distribution centered
at $\egamma^{0}$.  We perform Monte Carlo simulations using the method
presented in \cite{ldg2005}, and employing the detector thresholds from
the WXM on \hetetwonosp.

\begin{figure}[htb]
\begin{center}
\rotatebox{270}{\resizebox{3.5cm}{!}{\includegraphics{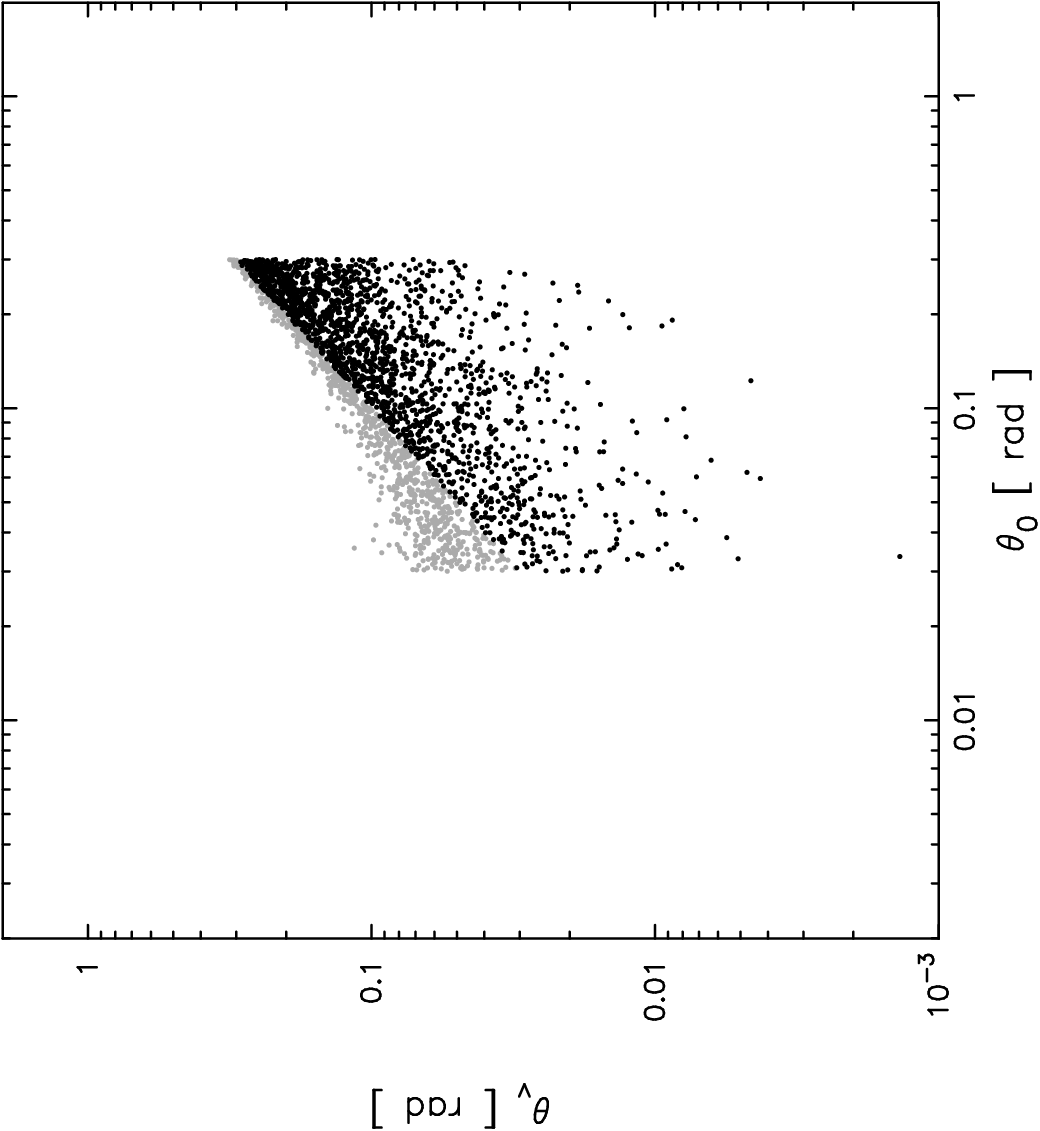}}}
\rotatebox{270}{\resizebox{3.5cm}{!}{\includegraphics{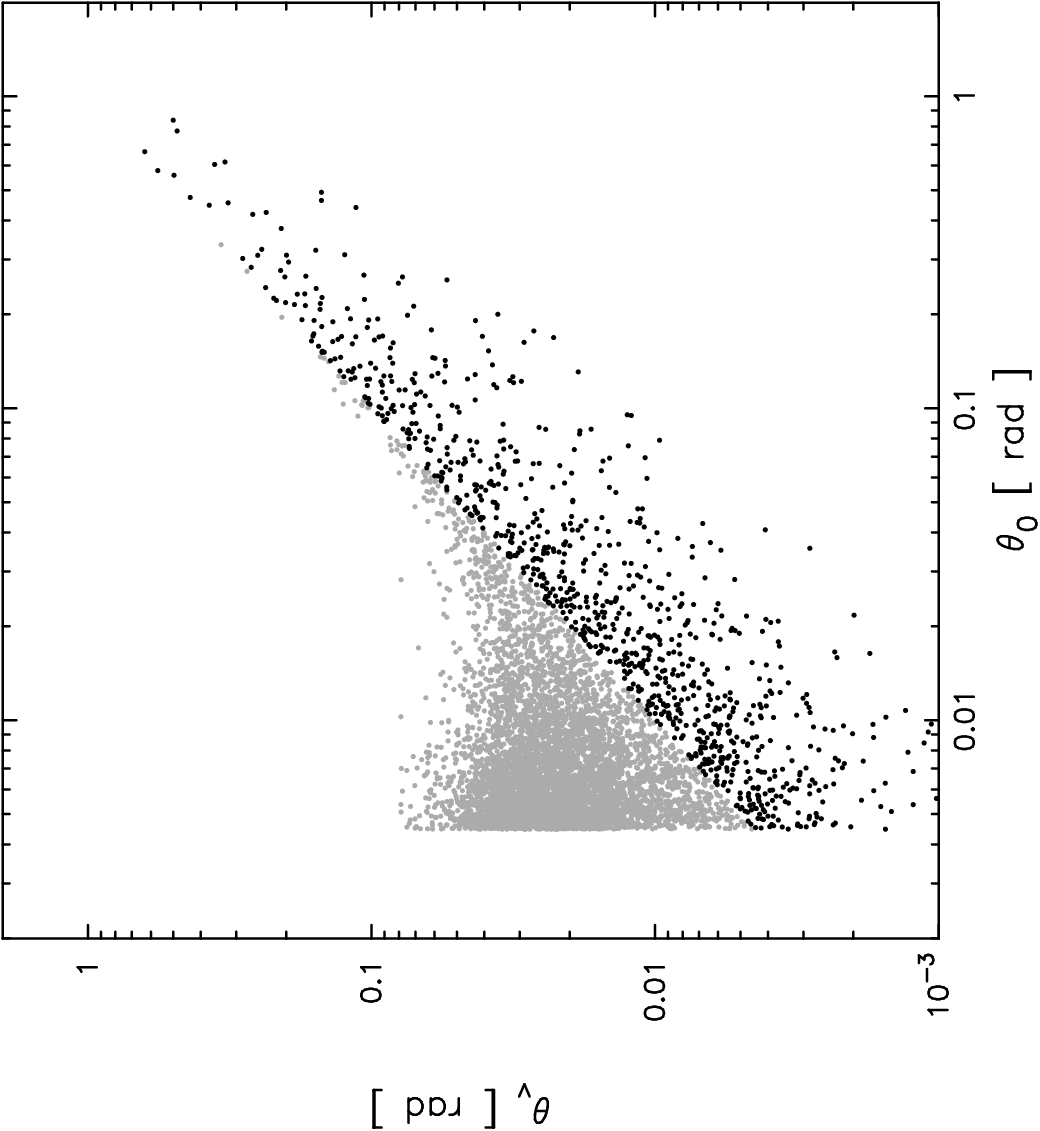}}}
\rotatebox{270}{\resizebox{3.5cm}{!}{\includegraphics{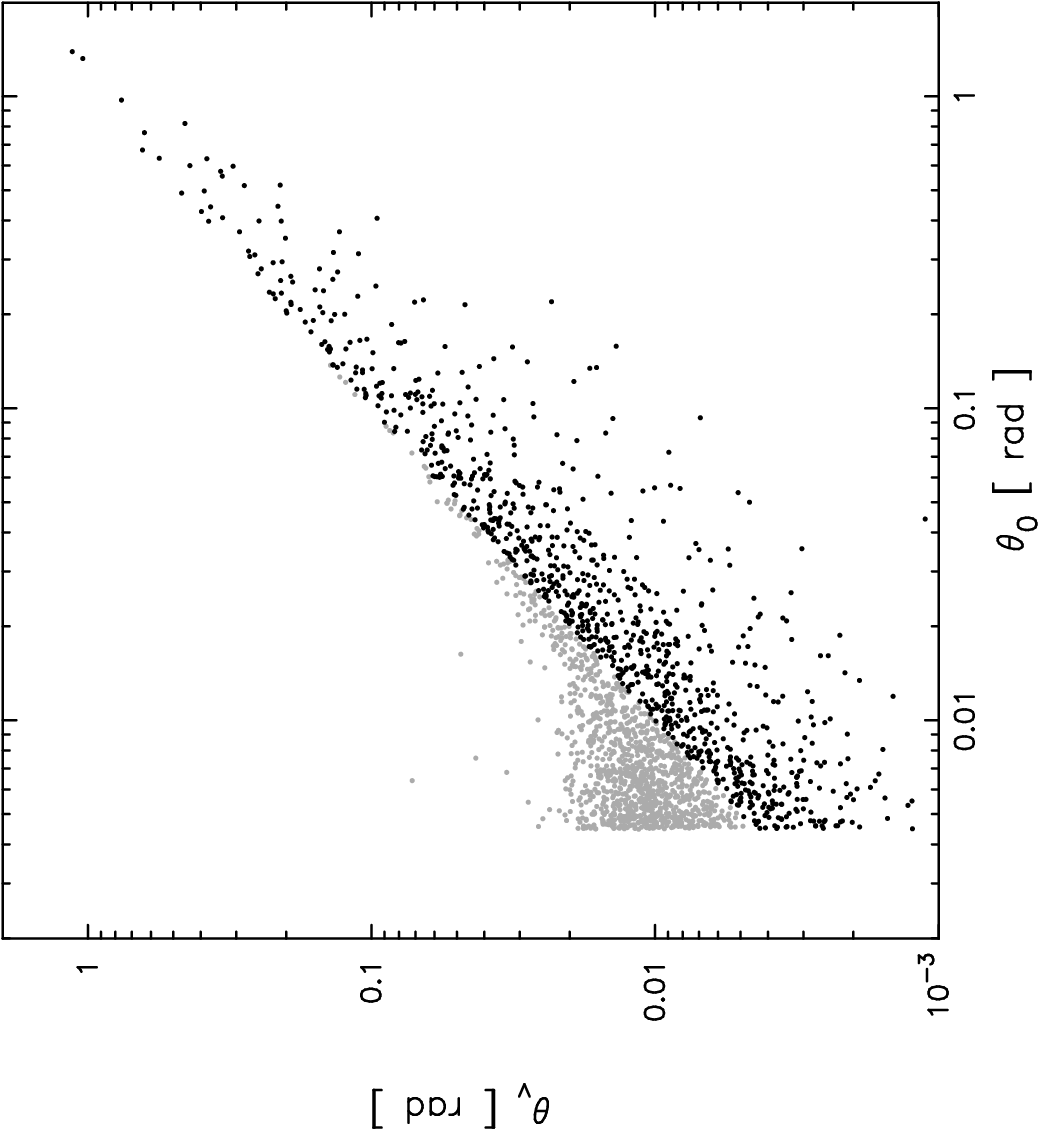}}}
\end{center}
\begin{center}
\rotatebox{270}{\resizebox{3.5cm}{!}{\includegraphics{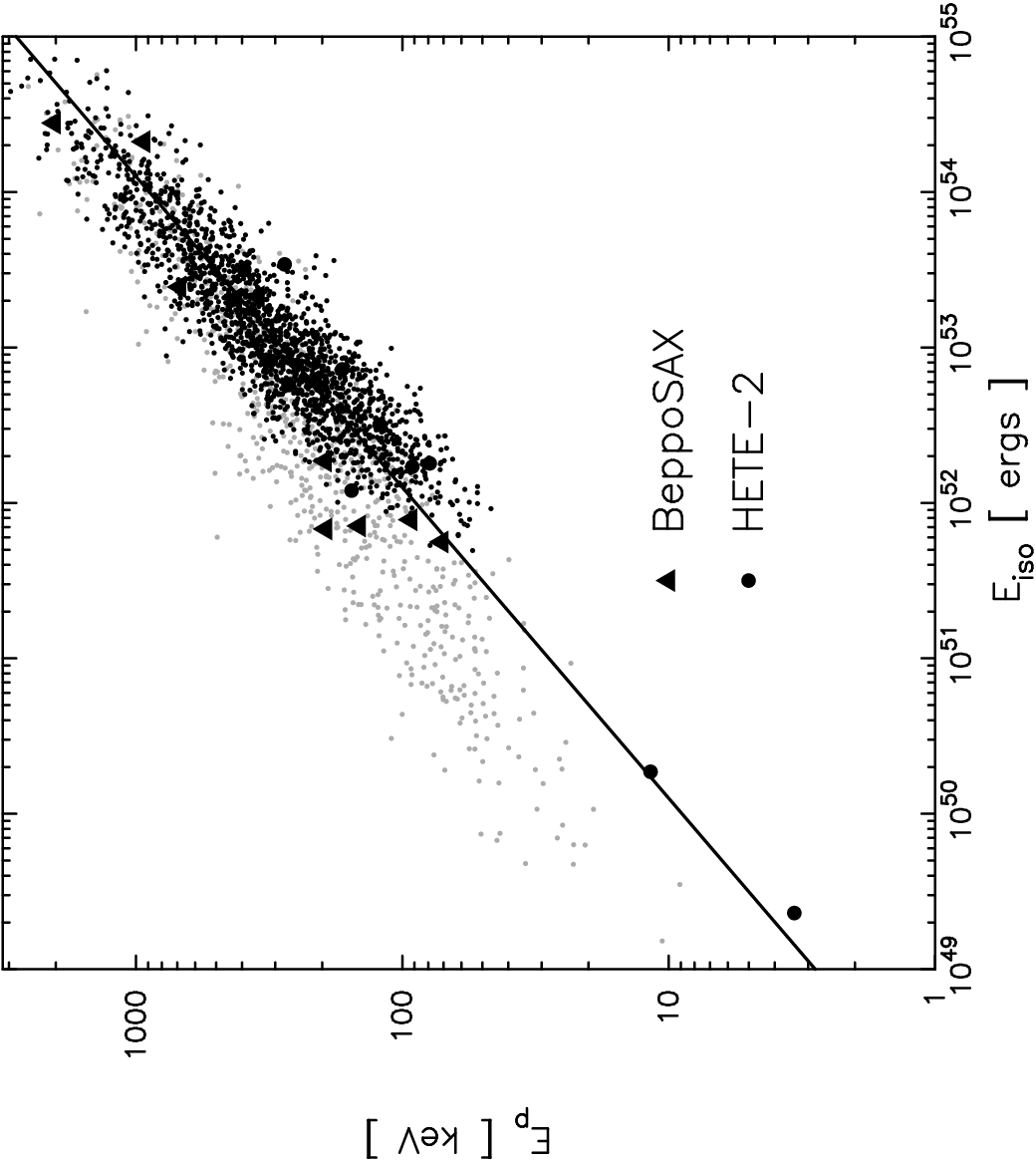}}}
\rotatebox{270}{\resizebox{3.5cm}{!}{\includegraphics{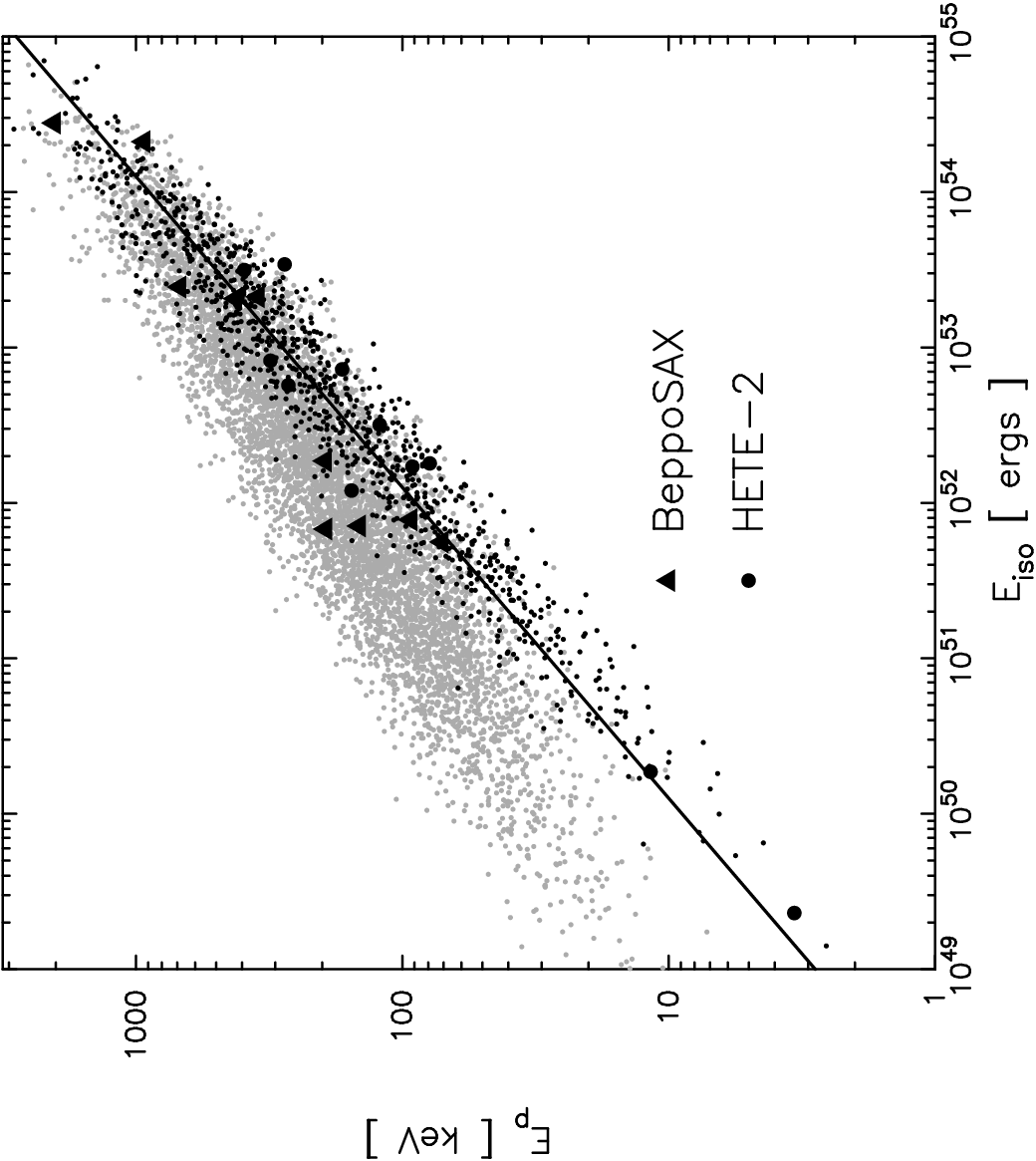}}}
\rotatebox{270}{\resizebox{3.5cm}{!}{\includegraphics{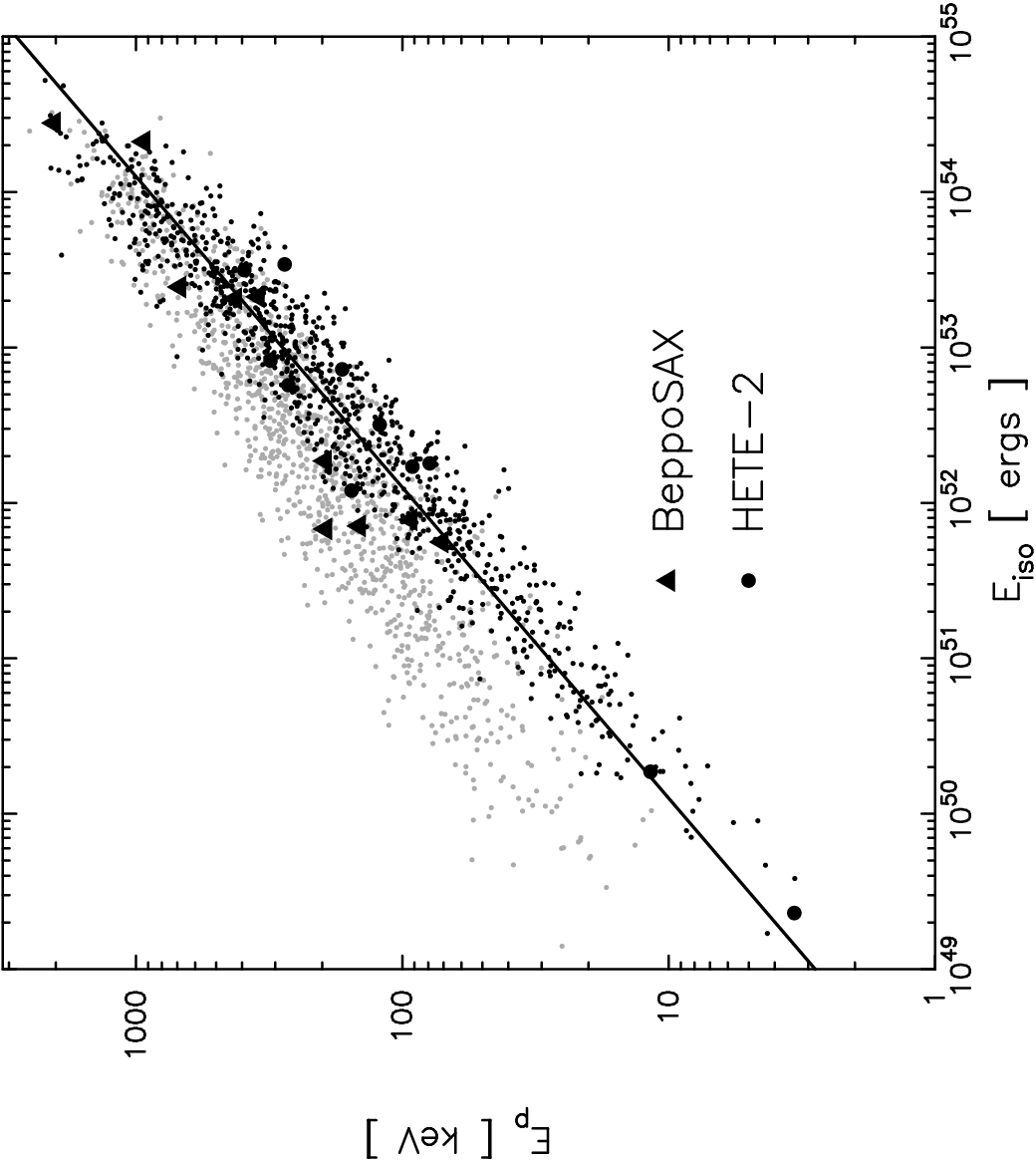}}}
\end{center}
\begin{center}
\rotatebox{270}{\resizebox{3.5cm}{!}{\includegraphics{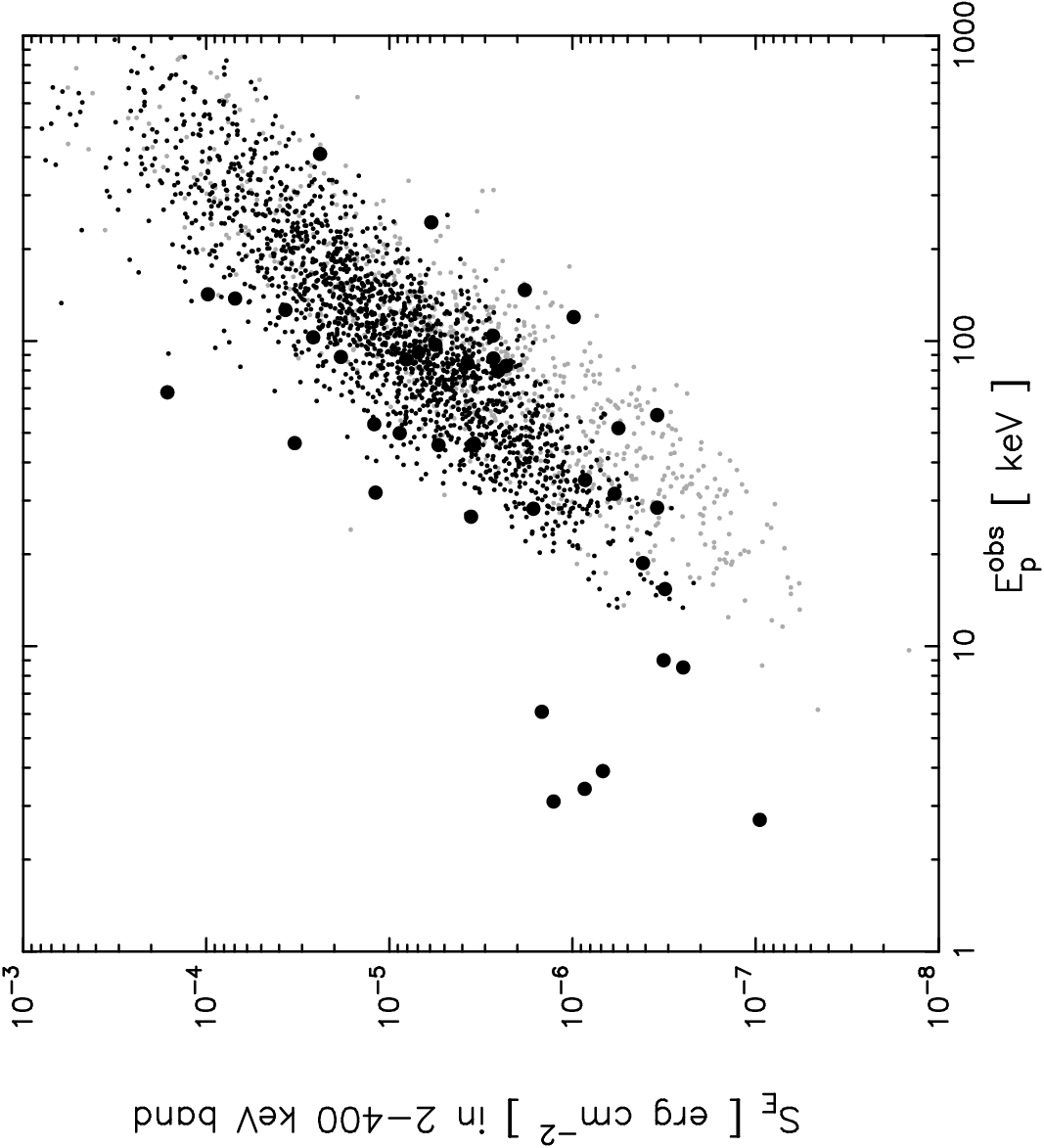}}}
\rotatebox{270}{\resizebox{3.5cm}{!}{\includegraphics{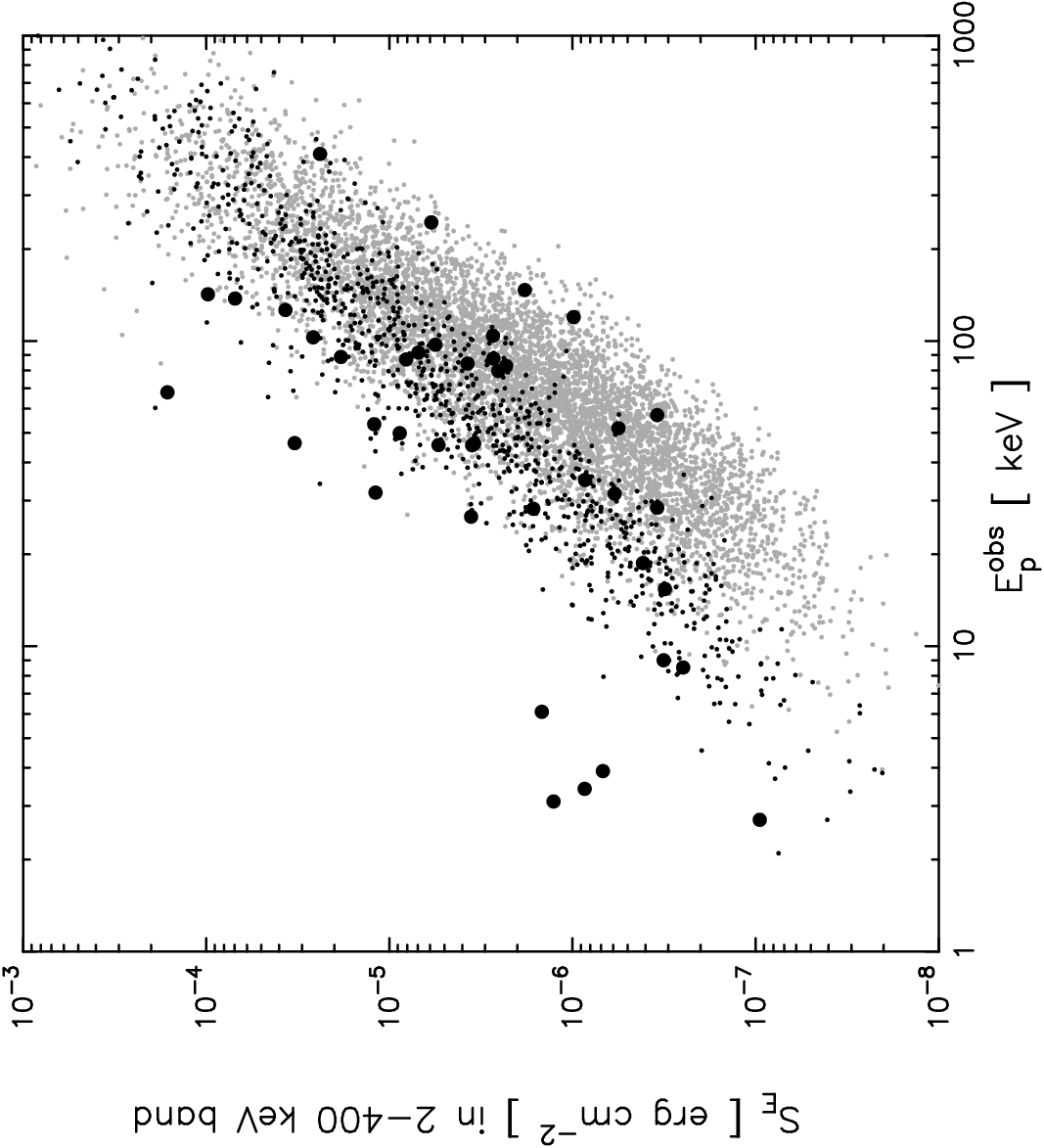}}}
\rotatebox{270}{\resizebox{3.5cm}{!}{\includegraphics{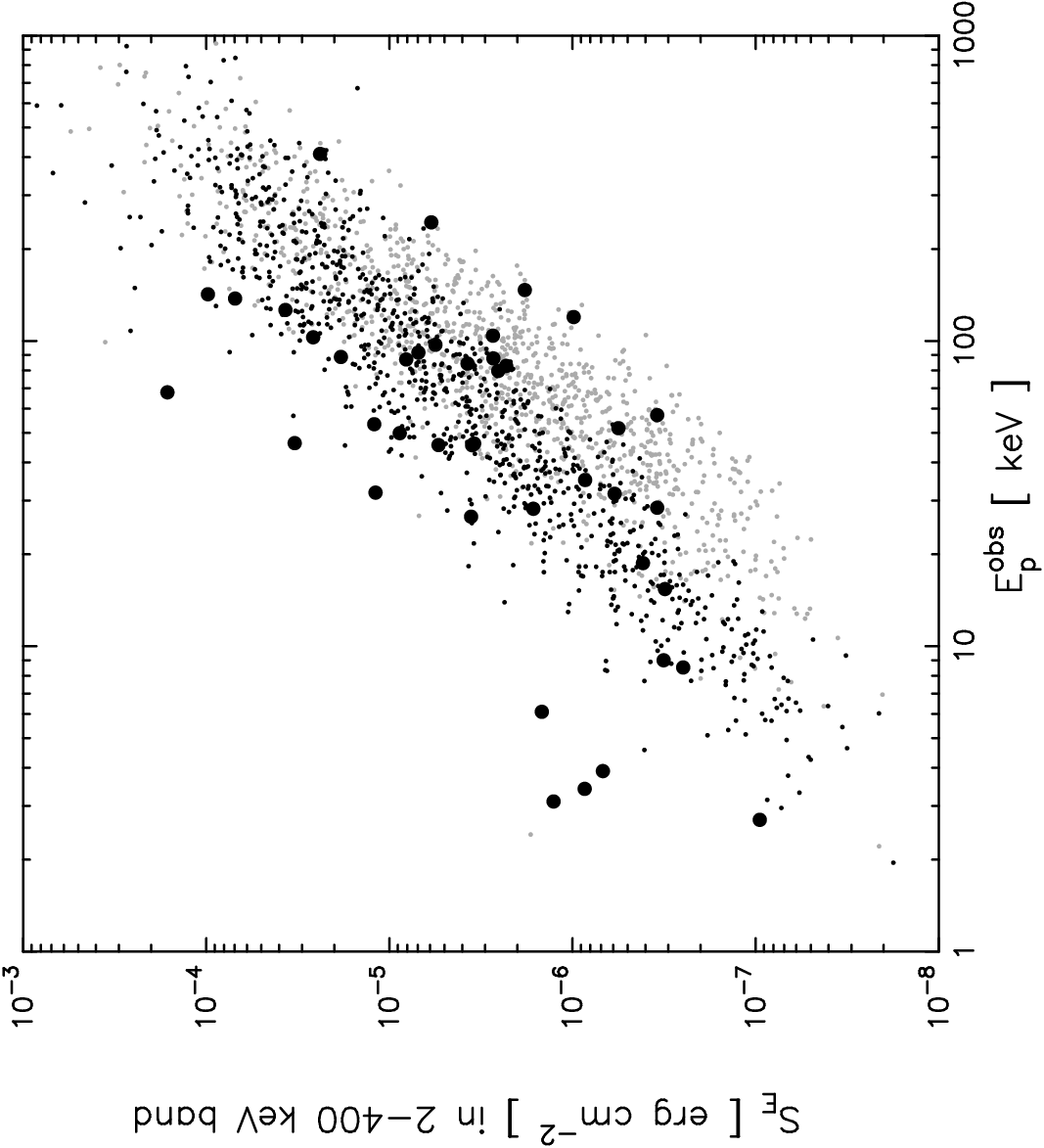}}}
\end{center}
\caption{Distribution of bursts detected on-axis (black) and off-axis
(gray) in the [$\thetaj$,$\thetav$]-plane (top row),
[$\eiso$,$\ep$]-plane (middle row) and [$\eop$,$S_{\rm
E}(2-400)$]-plane (bottom row) for the models Y04 (left), VOAUJ1
(center) and VOAUJ2 (right).  Bursts not detectable by the WXM are not
shown.}
\label{fig1}
\end{figure}

\section{Results}

Here we explore the relative importance of off-axis beaming for three
variable opening-angle uniform jet models.  The first (Y04) using the
parameters from \cite{yamazaki2004}, assumes $\gamma=100$ and draws
$\thetaj$ values from a power-law distribution given by
$f_{0}\,d\thetaj \propto \thetaj^{-2}\,d\thetaj$, defined from $0.3$ to
$0.03$ rad.  The Y04 model attempts to explain classical GRBs in terms
of the variation of jet opening-angles, while XRFs are interpreted as
bursts viewed off-axis.  The other two models explain both GRBs and
XRFs by a distribution of jet opening-angles, following results
presented in \cite{ldg2005}.  Here we add the presence of off-axis
beaming to this picture, considering both $\gamma=100$ (VOAUJ1) and
$\gamma=300$ (VOAUJ2).

As can be seen from the top row of Figure \ref{fig1}, the relative
importance of off-axis events increases for models with a population of
very small opening-angles.  This is mainly due to the fact that
narrower jets with a constant $\egamma$ will have larger $\eiso$
values, and therefore such bursts viewed off-axis will also be
brighter.  More importantly, the middle and bottom rows show that the
\hetetwo XRFs are not easily explained as classical GRBs viewed
off-axis.  The two XRFs with known redshift lie along the $\epei$
relation, and furthermore the larger sample of \hetetwo XRFs without
known redshifts do not fall in the region of the [$\eop$,$S_{\rm
E}$]-plane expected for this model; they lie at lower, rather than
higher, $\eop$ values for a given $S_{\rm E}$.  Even given the model of
the off-axis emission in \cite{yamazaki2004}, these \hetetwo XRFs are
difficult to explain.

The other two models we consider generate XRFs that obey the $\epei$
relation by extending the range of possible jet opening-angles to cover
five orders of magnitude (see \cite{ldg2005} for details and
discussion).  Hence, XRFs that obey the $\epei$ relation are bursts
that are seen on-axis, but have larger jet opening-angles. 
Nonetheless, these models generate a significant populations of
off-axis events, although increasing $\gamma$ reduces the fraction of
off-axis bursts in the observed sample.  

\section{Discussion}

Bursts with known redshift have been found to obey the $\epei$
relation, and a large population of off-axis bursts is not readily
apparent in the observed datasets.  \cite{liang2004} found that the
$\epei$ relation holds internally within a large sample of bright BATSE
bursts without redshift.  It is unknown whether fainter bursts might
deviate from this relation or what fraction of bursts are inconsistent
with the $\epei$ relation.  The result may be an indicator of the bulk
$\gamma$ of the material.

In future work we will investigate the effect of possible correlations
between $\thetaj$ and $\gamma$.  If narrower jets have larger bulk
$\gamma$ values, this could reduce the importance of off-axis beaming
even further.  Secondly, off-axis beaming will be important for
non-uniform jets as well \cite{graziani2005}.  Gaussian and
Fisher-shaped jets rely on the exponential falloff of the emissivity
with viewing angle to match the wide spread of observed burst
quantities \cite{zhang2004,dlg_rome1}.  If off-axis beaming is important,
the exponential falloff will be dominated at some angle by the
power-law falloff due to beaming, thereby broadening the emissivity
distribution.

\end{document}